\documentclass[twocolumn,superscriptaddress,showpacs,amsmath,amssymb]{revtex4}
\usepackage{graphicx}
\usepackage{latexsym}
\usepackage{color}
\usepackage{epsfig}

\begin{document}
\title{Acceleration of heavy and light particles in turbulence:
  comparison between experiments and direct numerical simulations} 

\author{R. Volk} \email{romain.volk@ens-lyon.fr}
\affiliation{Laboratoire de Physique, de l'\'Ecole normale
  sup\'erieure de Lyon, CNRS UMR5672, 46 All\'ee d'Italie, 69007 Lyon,
  France}
  
\author{E. Calzavarini} \affiliation{Faculty of Science, J.\ M.\
  Burgers Centre for Fluid Dynamics, and Impact-Institute, University
  of Twente, 7500 AE Enschede, The Netherlands}

\author{G. Verhille}\affiliation{Laboratoire de Physique, de l'\'Ecole normale
  sup\'erieure de Lyon, CNRS UMR5672, 46 All\'ee d'Italie, 69007 Lyon,
  France}

\author{D. Lohse} \affiliation{Faculty of Science, J.\ M.\
  Burgers Centre for Fluid Dynamics, and Impact-Institute, University
  of Twente, 7500 AE Enschede, The Netherlands}

\author{N. Mordant} \affiliation{Laboratoire de Physique Statistique
  de l'\'Ecole normale sup\'erieure de Paris, CNRS UMR8550, 24 rue
  Lhomond, 75005 Paris, France}

\author{J.-F. Pinton}\affiliation{Laboratoire de Physique, de l'\'Ecole normale
  sup\'erieure de Lyon, CNRS UMR5672, 46 All\'ee d'Italie, 69007 Lyon,
  France}

\author{F. Toschi} \affiliation{Istituto per le Applicazioni del
  Calcolo CNR, Viale del Policlinico 137, 00161 Roma, Italy\\ and
  INFN, Sezione di Ferrara, Via G. Saragat 1, I-44100 Ferrara, Italy.}

\begin{abstract}
  We compare experimental data and numerical simulations for the
  dynamics of inertial particles with finite density in turbulence. In
  the experiment, bubbles and solid particles are optically tracked in
  a turbulent flow of water using an Extended Laser Doppler
  Velocimetry technique. The probability density functions (PDF) of
  particle accelerations and their auto-correlation in time are
  computed. Numerical results are obtained from a direct numerical
  simulation in which a suspension of passive pointwise particles is
  tracked, with the same finite density and the same response time as
  in the experiment. We observe a good agreement for both the variance
  of acceleration and the autocorrelation timescale of the dynamics;
  small discrepancies on the shape of the acceleration PDF are
  observed. We discuss the effects induced by the finite size of the
  particles, not taken into account in the present numerical
  simulations.
\end{abstract}
\maketitle
\section{\label{intro}Introduction}
Understanding the transport of inertial particles with finite density,
such as sediments, neutrally buoyant particles or bubbles in turbulent
flows of water is of practical interest for both industrial
engineering or environmental problems. In a turbulent flow, the
mismatch in density between the particles and the fluid causes light
particles to be trapped in high vortical regions while heavy particles
are ejected form vortex cores and concentrate in high strain
regions~\cite{SundaramCollins}. As particles with different buoyancy
tend to concentrate in different regions of the flow, they are
expected to exhibit different dynamical behaviors. In recent years,
significant progress has been made in the limit of infinitely heavy,
pointwise particles~\cite{TheoInertial,cencini:2006}, and numerical
simulations have received experimental support~\cite{Eaton,Warhaft}.
In case of infinitely light particles (\textit{bubbles}): the result
of the numerical simulations on particle distributions and on fluid
velocity spectra \cite{wang93,maz03a,maz03b} agree in various aspects
with experimental findings \cite{ren05,ber06b,ber06c,cal07a} although
direct comparison between experiments and numerical simulations for
the acceleration pdf and correlation of the particles as not been
investigated in the past.

Indeed, in spite of the growing resolution of Direct Numerical
Simulations (DNS) of the Navier-Stokes equations at high Reynolds
numbers, it remains a challenge to resolve the motion of realistic
inertial particles: some degree of modelization is necessary. The
equation of motion of finite-size, finite-density particles moving in
a turbulent flow is not precisely known, and a comparison with
experimental data can help refining the models and extending their
range of validity.

Several experimental techniques have been developed for measuring the
velocity of particles along their trajectories. The optical tracking
method developed in the Cornell group has revealed that fluid
particles experience extremely intense accelerations~\cite{Voth},
while individual particles have been tracked for time durations of the
order of the flow integral time scale using an acoustic
technique~\cite{lyon1}. Because of the very fast decrease of the
acoustic scattering cross-section with the scatterer's size, this
method is limited to particles with diameter of the order of the
wavelength, {\it i.e.} inertial range
sizes~\cite{lyon2,bourgoin_legi}. The principle of the acoustic
technique is completely analogous to laser Doppler velocimetry (LDV),
provided that expanded light beams are used (an arrangement we call
E-LDV hereafter). The advantage of E-LDV, compared to acoustics, is
that the much smaller wavelength of light allows a better resolution
in space and also the use of smaller tracer particles.  The principle
of the measurement technique is reported in~\cite{volk_arxiv}, where
its performance has been compared and validated against the
silicon-strip tracking~\cite{Voth,MordantCornell} of neutrally buoyant
Lagrangian tracers.  We focus here on the dynamics of inertial
particles {\it i.e.}  particles whose density differ from that of the
fluid. Here we report the first comparison between experimental
measurements of acceleration of particles having a relative density in
the range $10^{-3}$ (air bubbles) to $1.4 $ (PMMA) in the same highly
turbulent flow, and numerical results obtained by tracking pointwise
particles with finite density in a direct numerical simulation of
isotropic homogeneous turbulence \cite{cal07b,cal07c}.

Numerical simulations are performed by means of standard
pseudo-spectral methods, where particular care has been used in
keeping a good resolution at the dissipative scales. The numerical
code for integrating the evolution of the Eulerian field and the
Lagrangian tracing of particles is the same as described in
\cite{maz03a,maz03b,roma-dns}. A thorough validation of the numerical
approach, included the Lagrangian evolution of the tracers has
recently been performed against experimental measurements
\cite{roma-goettingen}.  The numerical integration of tracers has,
with respect to experiments, the clear advantage of a uniform, well
controlled geometry and very large statistics; on the opposite, the
resolution can be limited to small Reynolds numbers.  For what
concerns the treatment of realistic particles, i.e. particles with a
density mismatch and a ``finite'' size, the best modelization to use
is not clear and one of the main goal of this manuscript is indeed to
compare state-of-the-art Lagrangian data against numerical results
from a current modelization.

\begin{figure}
  \includegraphics[width=8.0cm]{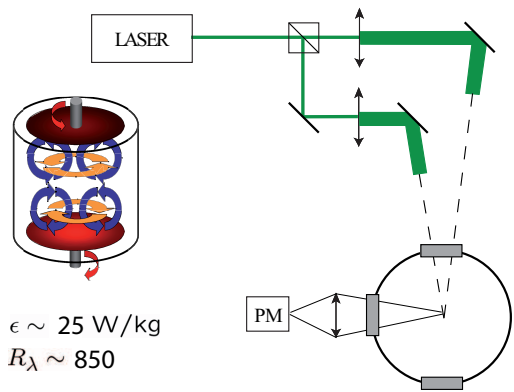}
  \caption{Experimental setup. (Top left): schematics of the von
    K\'arm\'an flow in water -- side view. (Top right): principle of
    the Laser-Doppler Velocimetry using wide beams (ELDV) -- top view
    of the experiment. PM: location of the photmultipler which detect
    scattering light modulation as a particle crosses the interference
    pattern created at the intersection of the laser beams.
  }\label{exp-setup}
\end{figure}

\section{\label{setup}Experimental setup and results}
The Laser Doppler technique is based on the same principle as the
ultrasound Doppler method which has good tracking performance of
individual Lagrangian tracers~\cite{lyon1,bourgoin_legi}.  In order to
access dissipative scales, and in particular for acceleration
measurements, one adapts the technique from ultrasound to laser light:
the gain is of a factor 1000 in wavelength so that one expects to
detect micron-sized particles. For a Lagrangian measurement, one has
to be able to follow the particle motion to get information about its
dynamics in time. For this, wide Laser beams are needed to illuminate
the particle on a significant fraction of its path. The optical setup
is an extension of the well known laser Doppler velocimetry technique;
Fig. \ref{exp-setup}. A Laser beam is split into two beams; each is
then expanded by a telescope so that their diameter is about 5mm. Then
the two beams intersect in the flow where they create an array of
interference fringes. As a particle crosses the fringes, the scattered
light is modulated at a frequency directly proportional to the
component of the velocity perpendicular to the fringes. It yields a
measurement of one component of the particle velocity. In practice, we
use a CW YAG laser of wave length 532 nm with 1.2 W maximum output
power. In order to get the sign of the velocity we use acousto-optic
modulators (AOM) to shift the frequency of the beams so that the
fringes are actually travelling at a constant speed. The angle of the
two beams is tuned to impose a 60 microns interfringe so that the
frequency shift between the beams (100~kHz) corresponds to 6~m/s. As
the beams are not focused, the interfringe remains constant across the
measurement volume whose size is about $5 \times 5 \times 10 \textrm{
  mm}^3$. It is imaged on a photomultiplier whose output is recorded
using a National Instrument PXI-NI5621 digitizer at rate $1$~MHz.

The flow is of the Von K\'arm\'an kind as in several previous
experiments using acoustics~\cite{lyon1} or optical
techniques~\cite{Voth}.  Water fills a cylindrical container of
internal diameter 15~cm, length~20 cm. It is driven by two disks of
diameter 10 cm, fitted with blades in order to increase steering. The
rotation rate is fixed at values up to 10~Hz. For the measurements
reported here, the Taylor based Reynolds number reaches up to 850 at a
maximum dissipation rate $\epsilon$ equal to 25 W/kg. We study three
types of particles: neutrally buoyant polystyrene particles with size
31 microns and density 1.06, PMMA particles with size 43 microns and
density 1.4 and air bubbles with a size of about 150 microns. The mean
size of the bubbles, measured optically by imaging the measurement
volume on a CCD, is imposed by the balance between the interfacial
surface tension $\sigma$ and the turbulent fluctuations of
pressure. This fragmentation process is known to lead to a well
defined and stationary size distribution \cite{lasheras} with a
typical diameter $D \propto
\left(\sigma/\rho_f\right)^{3/5}\epsilon^{-2/5}$, $\rho_f$ being the
density of the fluid.

The signal processing step is crucial as both time and frequency --
({\it i.e.}  velocity) -- resolutions rely on its
performance. Frequency demodulation is achieved using the same
algorithm as in the acoustic Doppler technique. It is a approximated
maximum likelihood method coupled with a Kalman filter~\cite{JASA}: a
parametric estimator assumes that the signal is made of a modulated
complex exponential and Gaussian noise. The amplitude of the recorded
signal and the modulation frequencies are assumed to be slowly
evolving compared to the duration of the time window used to estimate
the instantaneous frequency. Here the time window is about 30~$\mu$s
long and sets the time resolution of the algorithm. Outputs of the
algorithm are the instantaneous frequency, the amplitude of the
modulation and a confidence estimate which is used to eliminate
unreliable detections. Afterwards, the acceleration of the particle is
computed by differentiation of the velocity output. Note that
measurements are performed only when a particle moves within the
(limited) measurement volume so that after processing, the data
consists in a collection of sequences with variable lengths. For all
the measurements, the acceleration variance is computed using the same
procedure as in~\cite{MordantCornell}: it is obtained for several
width of the smoothing kernel used in the differentiation of the
velocity signal and then interpolated to zero filter width.

For small neutrally buoyant particles, {\it i.e.} Lagrangian tracers,
our data is in excellent agreement with the high speed imaging
measurements performed by the Cornell
group~\cite{Voth,MordantCornell,volk_arxiv}. When the rms value of the
acceleration is normalized by the Heisenberg-Yaglom scaling:
$\displaystyle \langle a^2 \rangle = a_0 \epsilon^{3/2}\nu^{-1/2}$
($\epsilon$ beeing the energy dissipation rate per unit mass and $\nu
= 1.3 \cdot 10^{-6}$~m$^2 $s$^{-1}$ the kinematic viscosity of
the fluid), both experiments yields the same values for the non
dimensional constant $a_0$ ($a_0 = 6.4 \pm 1$ at $R_\lambda=850$ for
the E-LDV compared to $6.2\pm 0.4$ for the Cornell data at $R_\lambda
= 690$).

\begin{table}
  \begin{center}
    \begin{tabular}{|l|c|c|c|c|c|}
   \hline
       \multicolumn{6}{|c|}{Experiment}\\
      \hline  
      Particle & radius $a$ & $\beta = \frac{3 \rho_f}{\rho_f + 2 \rho_p}$ & 
$St= \frac{\tau_p}{\tau_\eta}$ & $a_0$ & $a_{0}/a_{0,T}$\\

      \hline
      Tracers& $15.5~\mu$m & 0.96  & 0.24 & $6.4 \pm 1$ & $1$\\
      Neutral& $125~\mu$m & 0.96  & 16 & $2.2 \pm 1$ & $0.34$\\
      Heavy  & $20.5~\mu$m & 0.79  & 0.58 & $4.3 \pm 1$ & $0.67$\\
      Bubble & $75~\mu$m & 2.99  & 1.85 & $26 \pm 5$ & $4.06 $\\
      \hline
      \hline
     \multicolumn{6}{|c|}{Numerics}\\
        \hline  
        Particle & radius $a$ & $\beta = \frac{3 \rho_f}{\rho_f + 2 \rho_p}$ & $St= \frac{\tau_p}{\tau_\eta}$ & $a_0$ & $a_{0}/a_{0,T}$\\
      \hline
       Tracers & - &1  & 0.31 & $2.85 \pm 0.07$ & $1$ \\
      Neutral & - &1  & 4.1 & $2.94 \pm 0.07$ & $1.03$ \\
      Heavy   & - &0.75  & 1.03 & $2.63 \pm 0.12$ & $0.92$ \\
      Bubble  & - &3  & 1.64 & $25.9 \pm 0.46$ & $9.08$ \\
      \hline
    \end{tabular}
  \end{center} 
  \caption{(top) Parameters of the particles in the von
    K\'arm\'an flow at $R_\lambda=850$
    ($\eta=(\nu^3/\epsilon)^{1/4}=17~\mu$m and
    $\tau_\eta=\sqrt{\nu/\epsilon}=0.26~10^{-3}$~s). $\rho_p$ and
    $\rho_f$ are the densities of the particles and fluid, and
    $\tau_p=a^2/(3\beta \nu)$ is the stokes response time of the
    particles. The Taylor-based turbulent Reynolds number is
    computed as $R_\lambda=\sqrt{{15 u_{rms}^4}/{\epsilon\nu}}$
    measuring the one-component root-mean-square velocity, $u_{rms}$, with the E-LDV system and $\epsilon$ by
    monitoring the power consumption of the motors. The non
    dimensional constant $a_0$ is derived from the Heisenberg-Yaglom
    relationship. The last column compares the value for the inertial
    particle to the one obtained for the Lagrangian tracer (which is denoted by the subscript $T$).
    (bottom) Same as above: parameters of the particles tracked in the DNS
    of homogeneous isotropic turbulence at $R_\lambda=180$. 
    Out of the numerically analyzed 64 parameter combinations
    $(\beta , St)$, 
    we have picked those which were close to the experimental values for $(\beta , St)$.  }
\label{table1}
\end{table}
\begin{figure}
  \includegraphics[width=6.0cm,angle=-90]{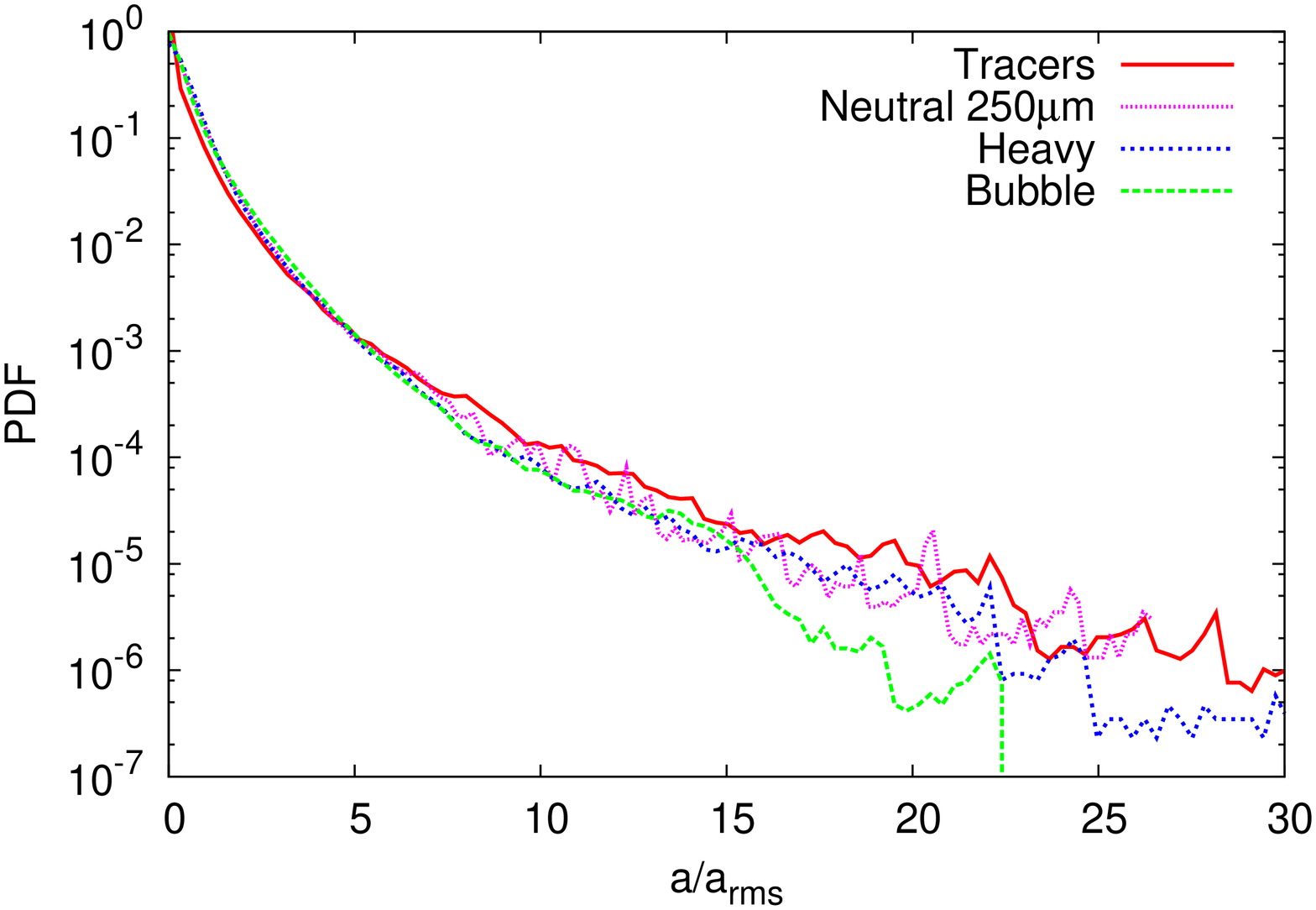}\\
  \includegraphics[width=6.0cm,angle=-90]{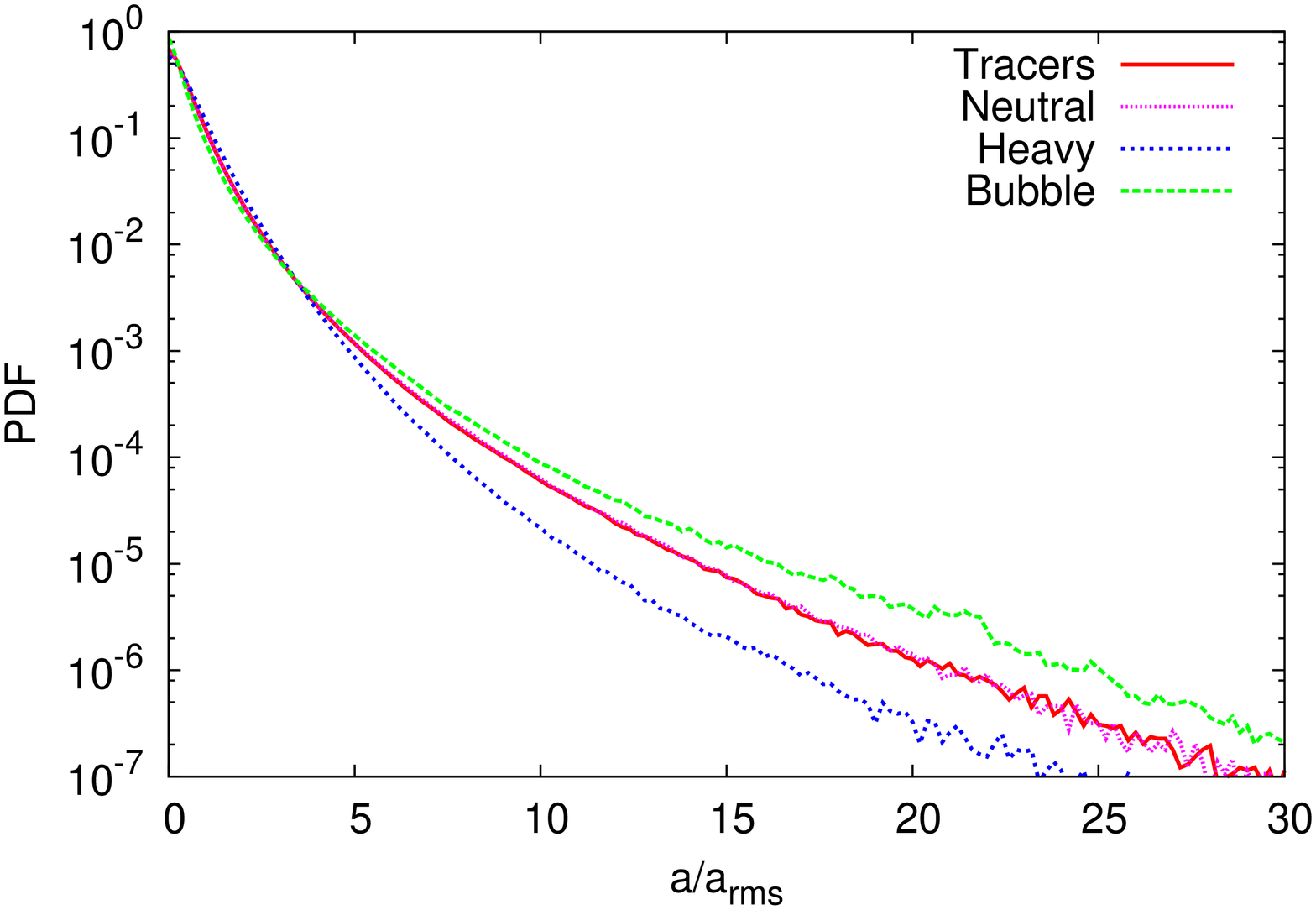}\\
  \includegraphics[width=6.0cm,angle=-90]{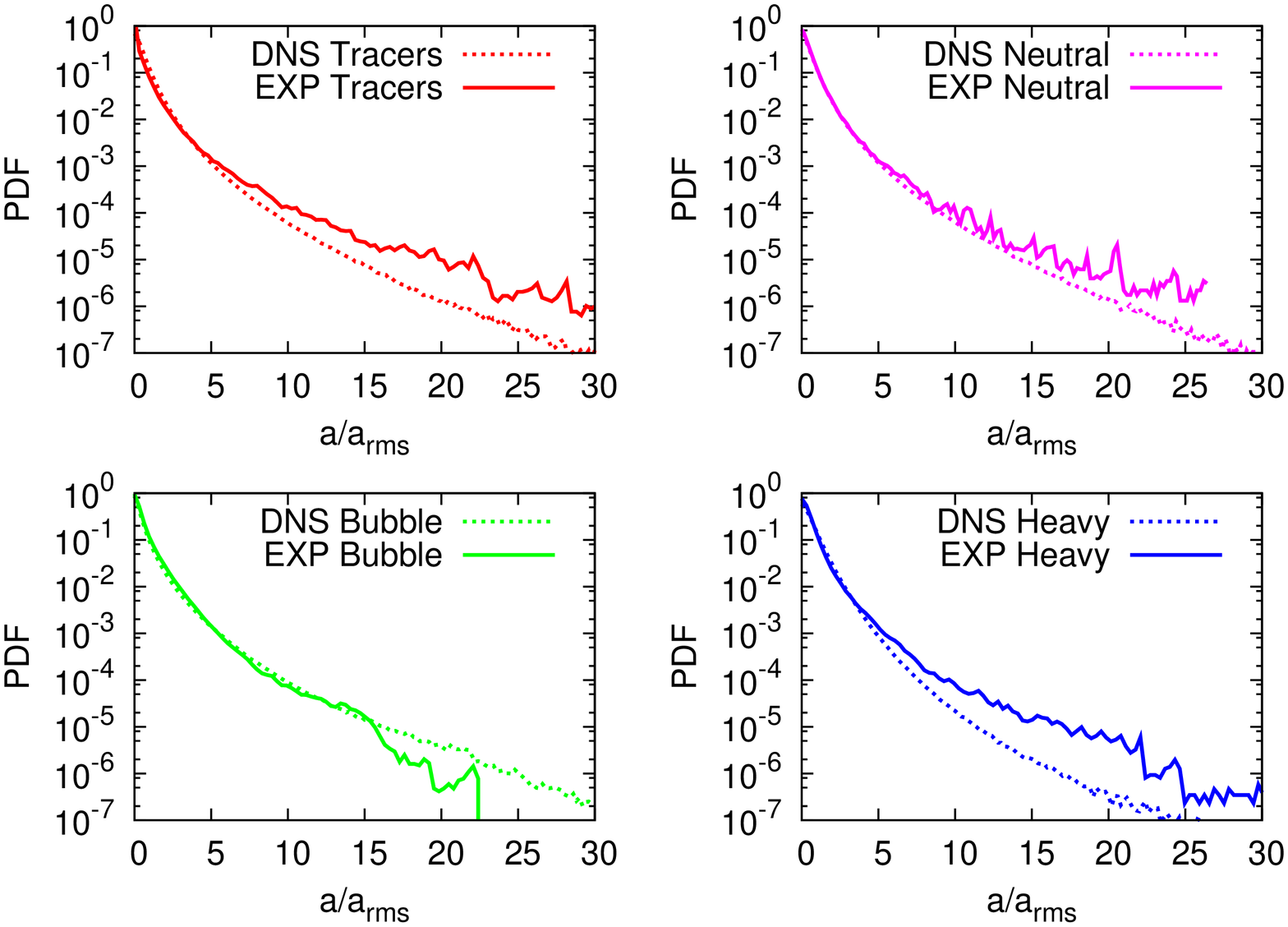}
  \caption{Probability distribution function of accelerations,
    normalized to the variance of the data sets.  (top) Data from
    experiment at $R_\lambda = 850$.  (middle) DNS of homogeneous
    isotropic turbulence at $R_\lambda = 180$.  (bottom) Comparison of
    experimental measurements and DNS results. }\label{fig-pdf}
\end{figure}
We have applied our technique to compare the dynamics of Lagrangian
tracers to the one of heavier or lighter particles (see table
\ref{table1} for numbers). We first compute the velocity root mean
square value $u_{rms}$ for the three cases: the values are $\{1.1,
1.2, 1.0\} \pm 0.1 \, {\rm m.s}^{-1}$ at $R_\lambda=850$ for the
tracers, heavy (PMMA spheres), and light particles (bubbles). Within
error bars, the large scale dynamics seems to be unaffected by changes
in the particle inertia. The acceleration distribution and
autocorrelation in the three cases are shown respectively in
Fig. \ref{fig-pdf} (top) and Fig. \ref{fig-cor} (top). The
acceleration PDFs are quite similar for moderate acceleration values
(below about $10 \, a_{\rm rms}$), as also observed in low Reynolds
number numerical simulations~\cite{Lohse}. However, the probability of
very large accelerations seems to be reduced in the case of inertial
particles as compared to Lagrangian tracers. The normalized
acceleration variance $a_0$ varies very significantly: it is reduced
to $4.3 \pm 1$ for heavier particles while it is increased to $26 \pm
5$ for bubbles. The correlation functions also show significant
changes with inertia: the characteristic time of decay is longer for
heavy particles and shorter for bubbles compared to tracers. We
measure $\tau_{\rm corr}/\tau_\eta = \{0.5, \; 0.9, \; 0.25\}$
respectively for tracers, heavy, and light particles, with the
correlation time defined as the half-width at mid amplitude of the
correlation function. We thus observe important changes in the
dynamics, even if the distribution of acceleration weakly changes with
inertia.

\indent Note that in our setup the Kolmogorov length is about $\eta =
17~\mu$m at $R_\lambda = 850$, so that the bubble size is about
$10~\eta$ and therefore may not be considered as a point
particle. Thus, one may wonder if the bubble dynamics is not altered
by spatial filtering as recently demonstrated for particles with
diameters in the inertia range~\cite{bourgoin_legi}. To check, we have
compared the dynamics of large neutrally buoyant particles with
diameter $250~\mu$m to the one of Lagrangian tracers. The results is
shown in Fig. \ref{fig-pdf} together with the other particles: the
effect of the particle size on the PDF is found to be weak as the
curve nicely superimposes with the ones for inertial
particles. However, the size effect is clear when comparing either the
coefficient $a_0$ (reduced to $2.2$), or the autocorrelation
functions.  One observes that the correlation time of the large
particles is twice the one for the tracers. We conclude that the
bubbles size may have a leading effect on the acceleration variance,
and that the value of $a_0$ reported here probably underestimates the
one that would be measured for smaller bubbles (with diameters closer
to the Kolmogorov scale).

\section{\label{numerics}Comparison with numerical simulations}
We compare the experimental data with the results from a direct
numerical simulation \cite{cal07b,cal07c} where a passive suspension
of pointwise particles with finite density are tracked in a
homogeneous isotropic turbulent flow.  The dynamics of the particles
is computed in the most simplified form of the equation of motion,
{\it i.e.} assuming that the particles are spherical, non-deformable
and smaller than the Kolmogorov length scale of the flow
~\cite{Maxey}.  When one retains only the Stokes drag force and the
added mass effect, the equation of motion then reads:
\begin{equation}
  \frac{d\mathbf v_p}{dt}=\beta \frac{D\mathbf u}{Dt}+\frac{1}{\tau_p}\left( \mathbf u - \mathbf v_p \right),
  \label{eqv}
\end{equation}
where $\mathbf v_p=\dot{x}(t)$ is the particle velocity, $\mathbf
u(x(t),t)$ the velocity of the fluid at the location of the particle
described by the Navier-Stokes equation, while $\beta = 3 \rho_f /
(\rho_f + 2 \rho_p)$ accounts for the added mass effect and and
$\tau_p = a^2/(3\beta \nu)$ is the Stokes response time for a particle
of radius $a$.  When made dimensionless by the Kolmogorov dissipative
scales ($\tau_{\eta}, \eta, u_{\eta}$) eq. (\ref{eqv}) reads
\begin{equation}
  {\mathbf a} \equiv \frac{d\mathbf v_p}{dt}=\beta \frac{D\mathbf u}{Dt}+\frac{1}{St}\left( \mathbf u - \mathbf v_p \right),
\end{equation}
with the particle acceleration ${\mathbf a}$ now expressed in the
Heisenberg-Yaglom units.  Thus, at a given Reynolds number, the
particles dynamics only depends on the values of the two dimensionless
parameters $\beta$ and $St=\tau_p/\tau_\eta$ This is generally
different from the case of infinite inertia of the particles
($\beta=0$), which has been formerly addressed in several numerical
and theoretical studies \cite{TheoInertial}, and for which instead
only the Stokes number $St$ matters. It is also different from the
pure bubble case ($\beta=3$) for which the particle indeed has no
inertia but only added mass \cite{wang93,maz03a,maz03b}.  We performed
numerical simulation at $Re_{\lambda} = 180$ (grid resolution
$512^3$), in which many particles, characterized by different pairs
$(\beta, St)$ (specifically 64 different sets of $O(10^5)$ particles)
were numerically integrated by means of eqn. (\ref{eqv}). Particles do
not have feedback on the flow field.

\begin{figure}
\begin{center}
  \includegraphics[width=6.0cm, angle= -90]{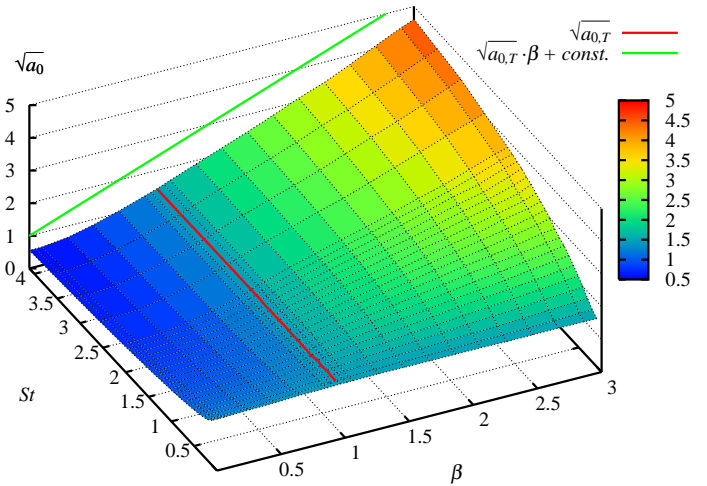}
  \caption{Behaviour of the normalized \textit{root mean square}
    acceleration $\sqrt{a_0} = ( \langle a^2 \rangle
    \epsilon^{-3/2}\nu^{1/2} )^{1/2} $ as a function of both $St$ and
    $\beta$ for a $Re_{\lambda}=75$ DNS.  Iso-contour for
    $\sqrt{a_{0,T}}$ (red) and the line $\sqrt{a_{0,T}} \cdot \beta +
    const.$ (green) are also reported.  Note that $a_0$ does not
    depend on $St$ for neutral ($\beta=1$) particles. While it is
    always reduced/enhanced for heavy/light particles. For large
    particles ($St \simeq 4.1$) we find $\sqrt{a_0} \simeq \beta
    \sqrt{a_{0,T}}$.}\label{a_rms_dns}
    \end{center}
\end{figure}

In order to compare the numerical results with the experimental data,
three types of particles (tracers, heavy, and bubbles) with different
inertia and Stokes number have been studied. The values for both
$\beta$ and $St$ have been chosen close to the ones of the particles
used for the E-LDV (see table \ref{table1}).  The evolution of the
normalized acceleration variance shows the same trend in experiments
and numerics: $a_0$ is reduced from the tracer value $2.85$ to $2.63$
for heavier particles and increased to $26$ for bubbles (table
\ref{table1}).  This seems to be a robust trend in the DNS.  To
emphasize this, in figure \ref{a_rms_dns} we show the behavior of
$\sqrt{a_0}$, i.e. the root-mean-square value of the particle
acceleration normalized by the Heisenberg-Yaglom scaling, in a wide
range of the $(\beta, St)$ parameter space from a less turbulent DNS
($Re_{\lambda} =75$) which has a very large number of $(\beta, St)$
pairs. Results from the $Re_{\lambda} =180$, not showed here, are
qualitatively similar. Note again that no significant Reynolds number
dependence of the probability distribution was found in ref.\
\cite{volk_arxiv}.

\begin{figure}
\includegraphics[width=6.0cm,angle=-90]{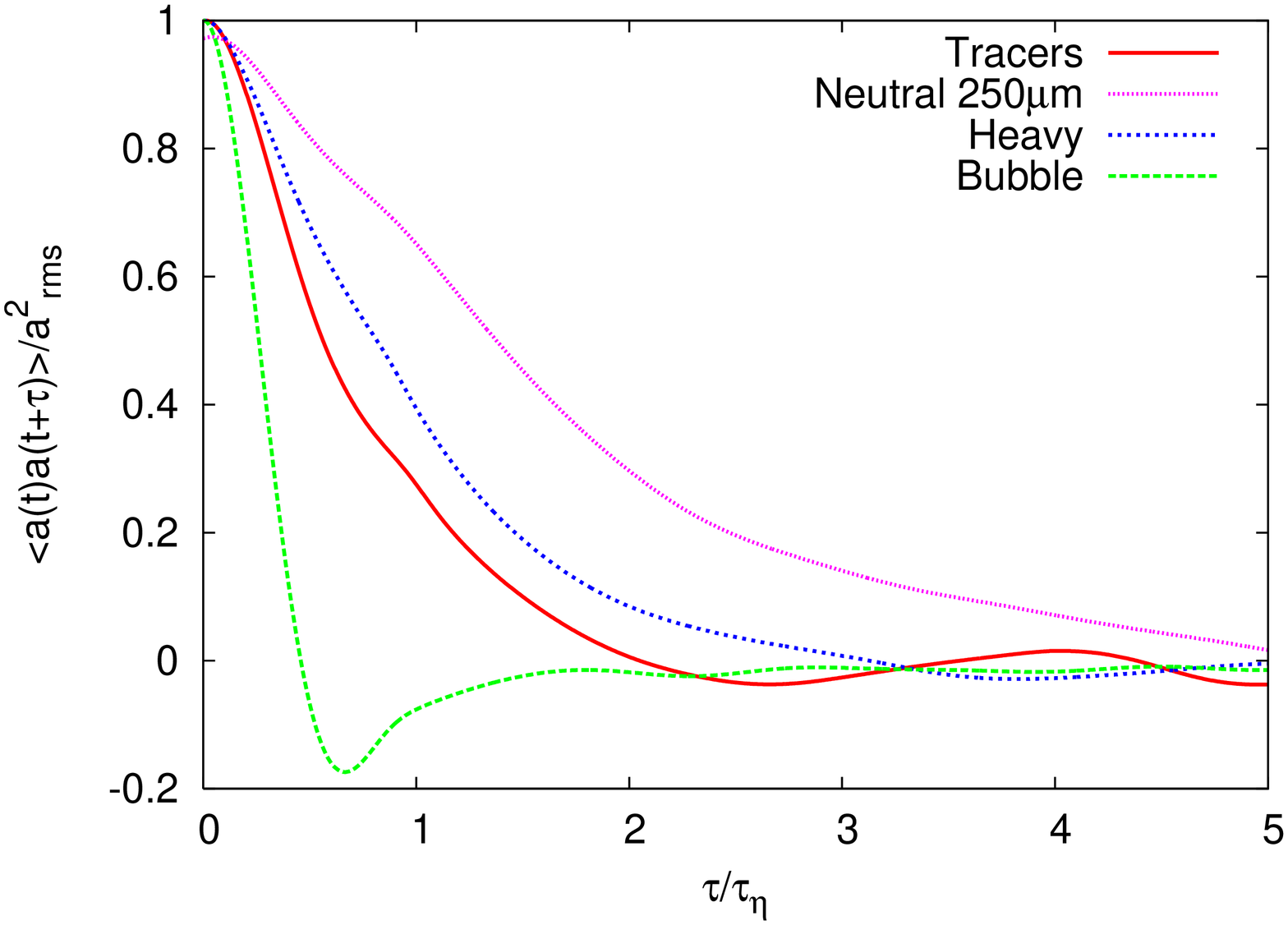}
\includegraphics[width=6.0cm,angle=-90]{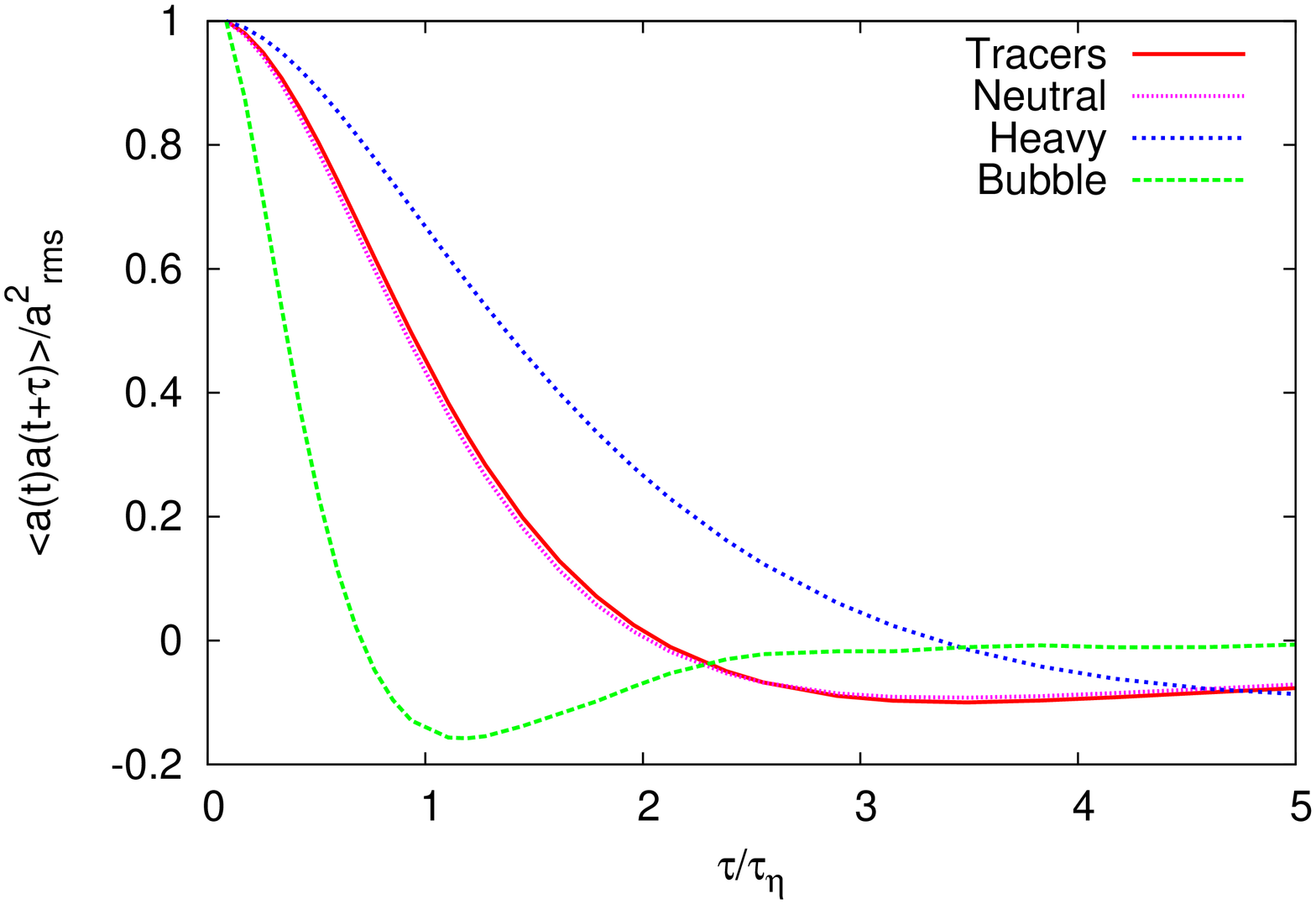}
\caption{Autocorrelation coefficients of the accelerations: (top)
  Data from experiments at $R_\lambda = 850$.  (bottom) Data from DNS
  of homogeneous isotropic turbulence at $R_\lambda = 180$ For the
  ($\beta,St$) values we refer to table \ref{table1}.
}\label{fig-cor}
\end{figure}

The acceleration distribution behavior and its comparison with the
experiment is reported in Fig. \ref{fig-pdf}.  In the numerics we
observe that the probability of very large accelerations is reduced
for the heavier particles as compared with tracers, while it is
increased for the bubbles. This feature,seems not to be present in the
experimental results.  Furthermore, we notice that for the three types
of particles, the acceleration PDFs, rescaled by the {\it rms}
acceleration, is close to the experiments. Experimental ones have
always longer tails, reflecting the more intermittent nature of the
turbulent flow, which has a larger Reynolds number ($Re_{\lambda,EXP}
= 850$ vs. $Re_{\lambda,DNS} = 180$ ).  We also observe a qualitative
agreement for the changes in the acceleration autocorrelation
functions when changing inertia, figure \ref{fig-cor}. One measures
$\tau_{\rm corr}/\tau_\eta = \{0.95, \; 1.35, \; 0.25\}$ respectively
for tracers, heavy and light particles. Just as observed for the
experiments, the dynamics is faster for the bubbles while heavier
particles decorrelate slower than fluid tracers. The $Re_{\lambda}$
difference is more pronounced here than in the pdf's (see ref.\
\cite{volk_arxiv}) and prevent from a more detailed comparison.

\section{\label{discussion}Discussion}
While solving a simplified version for the equation of motion, the
numerics reproduces qualitatively the effect of the particles inertia
on their dynamics. In particular, the dependence of the acceleration
autocorrelation on the particle inertia is nicely reproduced, see
figure \ref{fig-cor}.  However, also some discrepancies become
visible, though not yet completely conclusive, as a better resolution
and statistics of both the experiments and the numerics would be
important for firmer conclusions. Nevertheless, in this section we
shall have a closer look at the differences and propose some
explanations.

First of all, there is only qualitative agreement on the ratio
$a_{0,H}/a_{0,T}$. It is larger for the experiment than for the
numerics. Moreover, the tails of the numerical PDF of the bubble
acceleration seem to be enhanced as compared to those for tracer
acceleration. Vice versa, the tails of the numerical PDF of the
particle acceleration seem to be reduced as compared to those for
tracer acceleration.

What is the origin of the difference between the experiments and the
numerics?  First of all the Taylor-Reynolds numbers are different, but
ref.\ \cite{volk_arxiv} suggests an at most weak dependence of the
acceleration PDFs on the Reynolds number; a finding that is supported
by a comparison of our numerical simulations at $Re_\lambda = 185$ and
$Re_\lambda = 75$.

Next, in the numerical simulations we disregarded the lift and the
gravitational force. While this presumably has little effect on heavy
particles and tracer, it does modify the dynamics of the bubbles.  In
refs.\ \cite{maz03a,maz03b} we had shown by comparison of numerical
simulations for point bubbles with and without lift, that without lift
the bubble accumulation inside the vortices is more pronounced, i.e.,
bubbles without lift are more exposed to the small scale fluctuations,
which clearly will contribute to the pronounced tails of the
numerically found acceleration PDF, see figure \ref{fig-cor}, bottom.

Next, also the two-way coupling of the particles (i.e., the
back-reaction of the particles on the flow due to their buoyancy
difference) has been neglected in the simulations of this paper.  As
e.g.\ shown in refs.\ \cite{maz03a,maz03b} for bubbles and in ref.\
\cite{os} for particles, it has an effect on the turbulent energy
spectrum and thus also on the acceleration statistics.  However, as in
the present experiments the particle and bubble concentrations are
very low, the two-way coupling effect on the spectra should be hardly
detectable.

The final difference between numerics and experiments we will discuss
here -- and presumably the most relevant one -- is the finite size of
the particles in the experiments as compared to the numerics which is
based on effective forces on a point particle. Although the heavy
particles are not large as compared to $\eta$, this clearly holds for
the bubbles and the $250\mu m$ diameter neutral particles. Indeed,
figure \ref{fig-cor} shows how the finite size of these particles
smears out the acceleration autocorrelation, as compared to the tracer
case.  Also the ratio $a_{0,N}/a_{0,T}$ for large neutral particles is
only $0.34$, which demonstrates that the size of large particles has a
large effect on their acceleration variance. This type of spatial
filtering, which also lowers the PDF of large neutral particles in the
experiment, is not related to a temporal filtering of the particle
based on its response time. This is clearly visible in Fig.\
\ref{fig-pdf}(middle) where one can see that two neutral particles
($\beta=1$) with different response times (different $St$ or $\tau_p$)
have the same acceleration PDF, with same $a_0$, and same
autocorrelation function. Thus this size effect, which is not taken
into account in the point-particle based numerical simulations,
presumably is responsible for both the relatively small value of
$a_{0,B}/a_{0,T}$ measured for bubbles, and the change in the shape of
the PDF.

To conclude, we have reported acceleration measurements of inertial
particles using extended Laser Doppler velocimetry and have compared
the experimental data to DNS simulations of the motion of pointwise
particles with finite density. We have observed a qualitative
agreement between experiments and numerics in the shape of the PDF and
of the autocorrelation function.  We have given arguments for the
small discrepancies.  An experimental study of the motion of bubbles
with smaller sizes is needed for a better comparison with the
numerical simulations. Also numerical simulations keeping into account
the finite size of particles would presumably improve the agreement
between experiments and numerical data and detailed comparisions as
the one presented in this paper help to reveal the limitations of
point-particle models.  Obviously, going beyond point-particles is
extremely challenging in numerical simulations. A first step in this
direction has e.g.\ been taken by Prosperetti and coworkers with their
\textit{Physalis} method \cite{zhang05} which presently is extended
towards turbulent flows \cite{naso07}.

{\bf Acknowledgement} RV, GV, NM and JFP are indebted to Artem
Petrosian for his help in setting-up up the optics of the experiment,
and to Mickael Bourgoin for many fruitful discussions.  We are
grateful to Massimo Cencini, who contributed to the numerical study.
This work was partially funded by the Region Rh\^one-Alpes, under
Emergence Contract No.\ 0501551301, and by ANR contract \#192604.  We
thank the CASPUR (Rome-ITALY) and SARA (Amsterdam, The Netherlands)
for computing time and technical support.


\begin{thebibliography}{99}

\bibitem{SundaramCollins} S. Sundaram, L.R. Collins {\it J. Fluid
    Mech.} {\bf 379}, 105 (1999).
\bibitem{TheoInertial} J. Bec, M. Cencini and R. Hillerbrand, {\em
    Physica D} {\bf 226}, 11(2007); J.~Bec, L.~Biferale, M.~Cencini,
  A.~Lanotte, S.~Musacchio, and F.~Toschi, {\em Phys. Rev. Lett.},
  {\bf 98} 084502 (2007).
\bibitem{cencini:2006} M.~Cencini, J.~Bec, L.~Biferale, G.~Boffetta,
  A.~Celani, A.~S. Lanotte, S.~Musacchio, and F.~Toschi, {\em J. of
    Turb.}, {\bf 7}(36), 1--17 (2006).
\bibitem{Eaton} J. R. Fessler, J. D. Kulick, and J. K. Eaton, {\em
    Phys. Fluids} {\bf 6}, 3742-3749 (1994).
\bibitem{Warhaft} S. Ayyalasomayajula, A. Gylfason, L. R. Collins,
  E. Bodenschatz and Z. Warhaft, {\em Phys. Rev. Let.} {\bf 97},
  144507 (2006).
\bibitem{wang93} L. Wang and M. R. Maxey {\em App. Sci. Res.}, {\bf
    51}, 291--296 (1993).
\bibitem{maz03a} I. Mazzitelli, D. Lohse and F. Toschi, {\em
    Phys. Fluids}, {\bf 15}, L5-L8 (2003).
\bibitem{maz03b} I. Mazzitelli, D. Lohse and F. Toschi, {\em J. Fluid
    Mech.}, {\bf 488}, 283-313 (2003).
\bibitem{ren05} J. M. Rensen, S. Luther and D. Lohse, {\em J. Fluid
    Mech.}, {\bf 538}, 153--187 (2005).
\bibitem{ber06b} T. H. van den Berg, S. Luther, I. Mazzitelli,
  J. Rensen, F.  Toschi and D. Lohse, {\em J. of Turb.}, {\bf 7},
  1--12 (2006).
\bibitem{ber06c} T. H. van den Berg, S. Luther and D. Lohse, {\em
    Phys. Fluids}, {\bf 18}, 038103 (2006).
\bibitem{cal07a} E. Calzavarini, T. H. van den Berg, D. Lohse and
  F. Toschi, submitted to {\em Phys. Fluids}, arXiv:0607255 (2007).
\bibitem{Voth} G. A. Voth, A. La Porta, A. M. Crawford, J. Alexander
  and E. Bodenschatz, {\em J. Fluid Mech}, {\bf 469}, 121 (2002).
\bibitem{lyon1} N. Mordant et al. {\em Phys. Rev. Lett.}, {\bf
    87}(21), 214501, (2001); N. Mordant, P. Metz, O. Michel,
  J.-F. Pinton, {\em Rev. Sci. Instr.} {\bf 76}, 025105 (2005).
\bibitem{lyon2} N. Mordant, E. L\'ev\^eque and J.-F. Pinton, {\em New
    J. Phys.} {\bf 6}, 116 (2004).
\bibitem{volk_arxiv} R. Volk, N. Mordant, G. Verhille and
  J.-F. Pinton, {\em arXiv:0708.3350}.
\bibitem{MordantCornell} N. Mordant, A. M. Crawford and
  E. Bodenschatz, {\em Physica D} {\bf 193} 245 (2004); N. Mordant,
  A.M. Crawford, E. Bodenschatz, {\em Phys. Rev. Lett.} {\bf 94},
  024501 (2004).
\bibitem{cal07b} E. Calzavarini, M. Cencini, D. Lohse, F. Toschi, in
  {\em Advances in Turbulence XI} ISBN: 978-3-540-72603-6, 418-420
  (2007).
\bibitem{cal07c} E. Calzavarini, M. Cencini, D. Lohse, F. Toschi {\it
    Turbulence induced segregation of inertial particles} {in
    preparation}.
\bibitem{lasheras} C. Martinez-Baz\'a, J.-L. Monta\~n\'es and
  J. C. Lasheras, {\em J. Fluid Mech.} {\bf 401} 183-207 (1999)
\bibitem{JASA} N. Mordant, O. Michel and J.-F. Pinton, {\em
    J. Acoust. Soc. Am.}, {\bf 112}, 108-119 (2002).
\bibitem{Lohse} I. M. Mazzitelli, D. Lohse {\em New J. Phys.} {\bf 6}, 
203 (2004).
\bibitem{bourgoin_legi} N. Qureshi, M. Bourgoin, C. Baudet,
  A. Cartellier, Y. Gagne, to appear in {\em Phys. Rev. Lett.} {\em
    arXiv:0706.3042}.
\bibitem{Maxey} M.R. Maxey and J. Riley, {\em Phys. Fluids} {\bf 26},
  883 (1983).
\bibitem{roma-dns} L.~Biferale, G.~Boffetta, A.~Celani, B.~Devenish,
  A.~Lanotte, and F.~Toschi.  {\em Phys. Rev. Lett.}, {\bf 93}, 064502
  (2004).
\bibitem{roma-goettingen} L.~Biferale, E.~Bodenschatz, M.~Cencini,
  A.~S. Lanotte, N.~T. Ouellette, F.~Toschi and H.~Xu, {\em
    arXiv:0708.0311} (2007).
\bibitem{os} M. Boivin, O. Simonin, and K. D. Squires, {\em J. Fluid
    Mech.}, {\bf 375}, 235--263 (1998).
\bibitem{zhang05} Z. Zhang and A. Prosperetti, {\em J. Comput. Phys.},
  {\bf 210}, 292--324 (2005); J. Appl. Mech. - Transactions of the
  ASME 70, 64-74 (2003).
\bibitem{naso07} A. Naso and A. Prosperetti,   
 Proceedings of ICMF 2007, Leipzig (D) July 9-13, 2007.
\end{thebibliography}
\end{document}